\documentclass[conference]{IEEEtran}

\usepackage{graphicx}
\usepackage{wrapfig}
\usepackage{amsmath}
\usepackage{url}
\usepackage{booktabs}
\usepackage{topcapt}
\usepackage{times}
\usepackage{textcomp}
\usepackage{sidecap}
\usepackage[]{mcode}
\usepackage{setspace}
\usepackage{multicol}
\usepackage{moresize}
\usepackage[normalem]{ulem}
\usepackage{newclude}
\usepackage{wrapfig}
\usepackage{graphicx} % for pdf, bitmapped graphics files
\usepackage{epsfig} % for postscript graphics files
\usepackage{mathptmx} % assumes new font selection scheme installed
\usepackage{times} % assumes new font selection scheme installed
\usepackage{amsmath} % assumes amsmath package installed
\usepackage{amssymb}  % assumes amsmath package installed
\usepackage[]{algorithm2e}

\begin{document}

\title{D4M: Bringing Associative Arrays to Database Engines}

% \author{\IEEEauthorblockN{Vijay Gadepally, Jeremy Kepner}
% \IEEEauthorblockA{MIT Computer Science and Artificial Intelligence Laboratory, Cambridge, MA}
% \IEEEauthorblockA{\{vijayg, kepner\}@ csail.mit.edu}}

\author{\IEEEauthorblockN{Vijay Gadepally$^{\dagger \ddagger}$, Jeremy
    Kepner$^{\dagger \ddagger +}$, William
    Arcand$^{\dagger}$, David  Bestor$^{\dagger}$, Bill Bergeron$^{\dagger}$, Chansup
    Byun$^{\dagger}$, \\ Lauren Edwards$^{\dagger}$,
    Matthew Hubbell$^{\dagger}$, Peter Michaleas$^{\dagger}$, Julie Mullen$^{\dagger}$, Andrew Prout$^{\dagger}$,
    Antonio Rosa$^{\dagger}$, \\ Charles Yee$^{\dagger}$, Albert Reuther$^{\dagger}$}
\IEEEauthorblockA{$^\dagger$ MIT Lincoln Laboratory,   $^{\ddagger}$ MIT
  Computer Science and Artificial Intelligence Laboratory, $^+$ MIT
  Math Department}
}

% \author{Vijay Gadepally}
% \author{Jeremy Kepner}
% \author{William Arcand}
% \author{David Bestor}
% \author{Bill Bergeron}
% \author{Chansup Byun}
% \author{Lauren Edwards}
% \author{Matthew Hubbell}
% \author{Peter Michaleas}
% \author{Julie Mullen}
% \author{Andrew Prout}
% \author{Antonio Rosa}
% \author{Charles Yee}
% \author{Albert Reuther}
\maketitle

\let\thefootnote\relax\footnote{Vijay Gadepally is the corresponding
  author and can be reached at vijayg [at] mit.edu. \\
     This material is based upon work supported by the National Science
Foundation under Grant No. DMS-1312831. Any opinions, findings, and
conclusions or recommendations expressed in this material are those of
the author(s) and do not necessarily reflect the views of the National
Science Foundation.\\
  978-1-4799-6233-4/14/\$31.00 \textcopyright 2014 IEEE}

\begin{abstract}
%\boldmath
The ability to collect and analyze large amounts of data is a growing
problem within the scientific community. The growing gap between data
and users calls for innovative tools that address the challenges faced
by big data volume, velocity and variety.  Numerous tools exist that
allow users to store, query and index these massive quantities of
data. Each storage or database engine comes with the promise of
dealing with complex data. Scientists and engineers who wish to use
these systems often quickly find that there is no single technology
that offers a panacea to the complexity of information. When using
multiple technologies, however, there is significant trouble in
designing the movement of information between storage and database
engines to support an end-to-end application along with a steep
learning curve associated with learning the nuances of each underlying
technology. In this article, we
present the Dynamic Distributed Dimensional Data Model (D4M) as a potential
tool to unify database and storage engine operations. Previous articles on D4M have
showcased the ability of D4M to interact with the popular NoSQL
Accumulo database. Recently however, D4M now operates on
a variety of backend storage or database engines while providing a
federated look to the end user through the use of associative
arrays. In order to showcase how new databases may be supported by
D4M, we describe the process of building the D4M-SciDB connector and
present performance of this connection.

\end{abstract}

\section{Introduction}

The challenges associated with big data are commonly referred to as
the 3 V's of Big Data - Volume, Velocity and
Variety~\cite{laney20013d}. The 3 V's provide a guide to the largest outstanding challenges associated with
working with big data systems. Big data \textbf{volume} stresses the storage,
memory and compute capacity of a computing system and requires access
to a computing cloud. The \textbf{velocity} of big data stresses the rate at which data can
be absorbed and meaningful answers produced. Big data \textbf{variety}
makes it difficult to develop algorithms and tools that can address
that large variety of input data.

The ability to collect and analyze large amounts of data is a growing
problem within the scientific community. The growing gap between data
and users calls for innovative tools that address the challenges faced
by big data volume, velocity and variety.  Numerous tools exist that
allow users to store, query and index these massive quantities of
data. Each storage or database engine comes with the promise of
dealing with complex data. Scientists and engineers who wish to use
these systems often quickly find that there is no single technology
that offers a panacea to the complexity of
information~\cite{stonebraker2005one, cattell2011scalable}. When using
multiple technologies, however, there is significant trouble in
designing the movement of information between storage and database
engines to support an end-to-end application. In this article, we
present the Dynamic Distributed Dimensional Data Model - a technology
developed at MIT Lincoln Laboratory. Previous articles on D4M~\cite{kepner2013d4m,kepner2012dynamic} have
showcased the ability of D4M to interact with the popular Apache
Accumulo database. Recent advances in D4M now allow D4M to operate on
a variety of back end storage or database engines while providing a
federated look to the end user through the use of associative
arrays. Associative arrays provide a mathematical interface across
different database technologies and can help solve one of the largest
problems of working with numerous backend storage or database engines
- how do we correlate information that may be spread across different
storage or database engines?

The Intel Science and Technology Center (ISTC) on Big Data~\cite{istcwebpage} is centered
at the MIT Lincoln Laboratory and supports five major research themes:
Big Data Databases and Analytics, Big Data Math and Algorithms, Big
Data Visualization, Big Data Architectures, and Streaming Big Data. One of the core goals of the ISTC is to develop the next generation software
stack required to manage heterogenous data in order to enable large
scale data analytics on data from the Internet of Things (IoT). This
solution stack is known as the Big Data Working Group (BigDAWG)
stack~\cite{dugganbigdawg}. The BigDAWG solution stack is a vertically integrated stack
that supports numerous hardware platforms, analytics libraries,
database and storage engines, software development through the Big
Dawg Query Language (BQL) and Compiler, visualization and presentation
of data through a variety of applications. The BQL will provide software and
analytics developers an abstraction of the underlying database and
storage engines, analytics libraries and hardware platforms. A key
feature of BQL is to develop the API required to provide a federated
look to developers.

Federated databases have the ability to abstract away details about
the underlying storage or database engine. Very often, federated
databases are used to provide some mutual benefit. This
feature can be quite appealing to scientists who wish to write complex
analytics and are not necessarily database or storage experts. There
has been much promise of federated
databases~\cite{sheth1990federated}. Federated databases provide the
ability to give users the feel of a data warehouse without physically
moving data into a central repository~\cite{haas2002data}. As an
example of a federated database, consider
Myria~\cite{halperin2014demonstration, howe2013collaborative}, a distributed database that uses
SQL or MyriaL as the language all of which was developed at the University of
Washington. One of the challenges in database federation has
been in developing a programming API that can be used to interact
with the ever-increasing variety of databases and storage
engines~\cite{cheneyprogramming}. 

D4M's mathematical foundation, associative arrays, have the ability to
to help alleviate the challenges associated with open
problems in federated database. Having a one-to-one relationship with
triple store or with key-value store systems allows a flexible
representation that can be supported by many databases. The ability to
perform linear algebraic operations on associative arrays (and thus
data stored in different database engines) opens up
big-data analytics to non-computer scientists. We believe that an API based on
mathematical operations is easy to learn. The software
implementation in popular languages such as MATLAB,
Octave, and Julia allows the rapid prototyping of new and complex
analytics with minimal effort.

In this article, we present our work on developing associative arrays
as the datatype for big data in Section~\ref{assoc}. In
Section~\ref{d4m}, we present D4M and provide examples of how database
operations such as context and cast can be done with D4M and
associative arrays through the D4M MATLAB toolbox. In Section~\ref{scidb}, in order to motivate the
ease at which new database support can be built into D4M, we detail
the D4M-SciDB connector.  In order to demonstrate the use of D4M, associative arrays, and
database engines, we provide a
simple case study of developing an analytic for medical data that
spans across three different storage engines in
Section~\ref{medical}. Finally, we conclude in
Section~\ref{conc}.

\section{Associative Arrays}
\label{assoc}

Associative arrays are
used to describe the relationship between multidimensional entities
using numeric/string keys and numeric/string values. Associative
arrays provide a generalization of sparse matrices. Formally, an
associative array \textbf{A} is a map from $d$ sets of keys $K_1 \times K_2 \times ... \times K_d$ to a value set $V$ with a semi-ring structure
$$
{\bf A}: K_1 \times K_2 \ ... \times K_d \rightarrow V
$$
where $(V,\oplus,\otimes, 0, 1)$ is a semi-ring with addition operator
$\oplus$, multiplication operator $\otimes$,
additive-identity/multiplicative-annihilator 0, and
multiplicative-identity 1.  Furthermore, associative arrays have a
finite number of non-zero values which means their support $supp({\bf
  A})={\bf A}^{-1} (V \backslash \{ 0 \} )$ is finite. While
associative arrays can be any number of dimensions, a common technique
to use associative arrays in databases is to project the d-dimensional
set into two dimensions as in:
$$
{\bf A}: K_1 \times \{ K_2 \cup K_3 \cup ... \cup K_d \} \rightarrow V
$$
where the $\cup$ operation indicates a union operation. In this 2D
representation, $K_1$ is often referred to as the row key and $ \{ K_2
\cup K_3 ... \cup K_d \} $ is referred to as the column key.

As a data structure, associative arrays return a value given some
number of keys and constitute a function between a set of tuples and a
value space. In practice, every associative array can be created from
an empty associative array by simply adding and subtracting values. With
this definition, it is assumed that only a finite number of tuples
will have values, and all other tuples have a default
value of the additive-identity/multiplicative-annihilator 0. Further, the associative array mapping should support
operations that resemble operations on ordinary vectors and matrices
such as matrix multiplication. In practice, associative arrays support a variety of linear algebraic operations
such as summation, union, intersection, multiplication and element
wise operations. Summation of
two associative arrays, for example, that do not have any common row
or column key performs a union of their underlying non-zero
keys. Element wise multiplication as an example performs an operation
similar to an intersection. Associative arrays have a one-to-one
relationship with key-value store or triple store databases, sparse
matrices, and adjacency or incidence matrix representations of
graphs. These relations allow complex datasets to be easily converted
to associative array representation. Linear algebraic operations on
associative arrays can be used to perform graph algorithms as described in~\cite{gabb2015}.

NoSQL database tables can be exactly described using the mathematics of
associative arrays~\cite{kepner2015associative}. In the D4M schema, a table in
a NoSQL database, such as Apache Accumulo, is an associative array. In
this context, the primary differences between associative arrays and
sparse matrices are: associative array entries always carry their
global row and column labels while sparse matrices do not. Another
difference between associative arrays is that sparse
matrices can have empty rows or columns while associative arrays do
not. 

Using associative arrays as a datatype for big data has many
benefits such as:

\begin{itemize}
\item Using associative arrays as the base datatype will make database
  development easier. DB developers will only need to provide an
  optimized interface to associative arrays;
\item Associative arrays can limit the programming language-DB connectors that are
  required. Currently, if there are N programming languages and M
  database engines, we need $N \times M$ connectors. Having a single
  interface can reduce this to $N + M$ connectors; and
\item An API based on mathematical operations is natural for the vast
  majority of scientists and engineers.
\end{itemize}

\section{The Dynamic Distributed Dimensional Data Model (D4M)}
\label{d4m}

The Dynamic Distributed Dimensional Data Model (D4M) combines techniques
from diverse fields to support rapid prototyping of big data
problems in a familiar programming environment. Specifically, D4M consists of 3 components:

\begin{itemize}
\item A software API that enables D4M to connect with databases,
\item A software API that supports Associative Arrays and their
  mathematics, and
\item A schema to represent unstructured multi-dimensional datasets.
\end{itemize}

\begin{figure}
\centerline{
\includegraphics[width=3.1in]{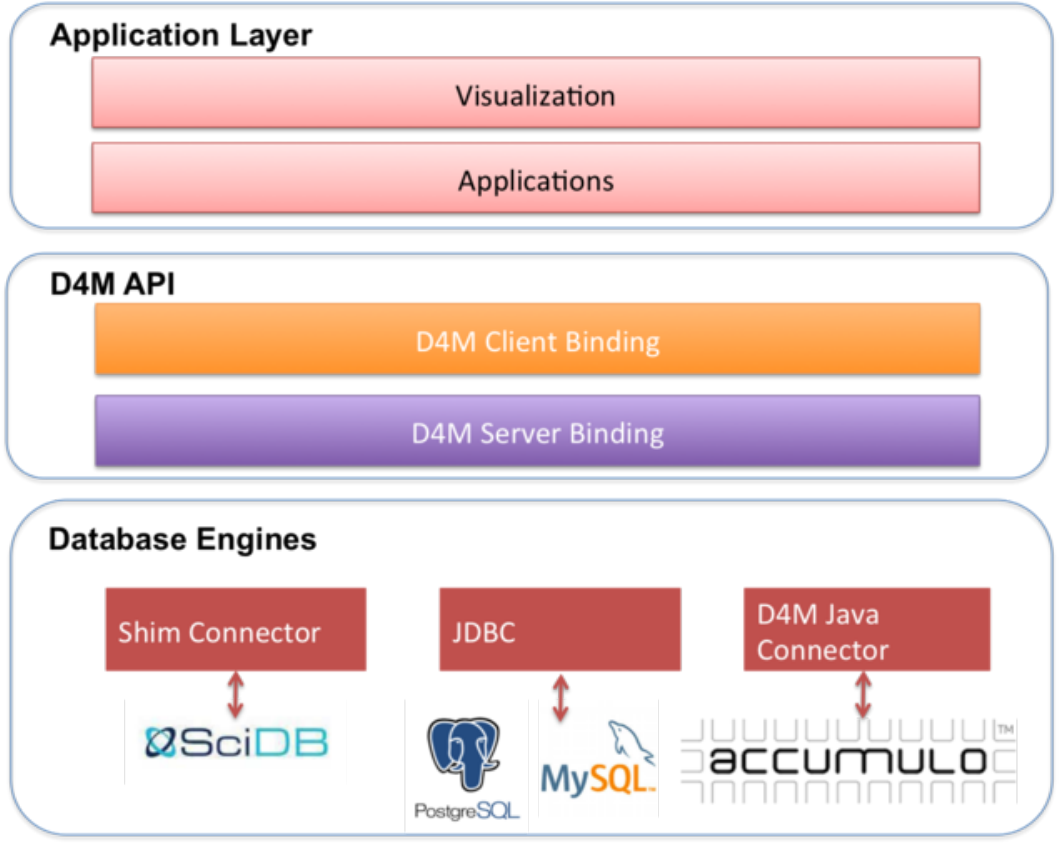}
}
\caption{D4M provides a middle layer between storage/database engines
  and applications. The D4M client binding provides support for
  associative arrays. The D4M server binding connects client code to
  different database engines.}
\label{d4marchitecture}
\end{figure}

D4M has a multi layer architecture that allows users to develop
analytics without knowledge of the underlying engine. In
Figure~\ref{d4marchitecture} we describe the various components of
D4M. The D4M software API is roughly broken into two components - a client
binding and a server binding. 

The D4M client binding is responsible for most of the
sophisticated data processing. Support for the associative array
datatype allows users to quickly convert a representative subset of
their dataset into an
associative array and prototype different algorithms to test for
mathematical correctness. Given the relationship between associative
arrays and sparse matrices, there are a wide variety of potentially
complex algorithms such as machine learning that can be directly
translated to operations on associative arrays. Once algorithms have
been developed and tested for correctness, a user can make use of the
D4M server binding to scale their dataset by connecting to a database engine. 

The D4M server binding allows users to map their in-memory associative
arrays to a wide variety of backend storage or database
engines. Creating a database server binding creates an object in local memory
that contains information about the database type, authentication
information, and host. Using this object, one can create a table object
that binds the D4M client to a DB table. With minimal effort, a user can read in raw data, convert to associative
array representation, and insert into a database. Querying from the
database results in associative arrays that can be directly used for
the complex analytics developed using the client binding. Syntax wise,
querying data from an associative array or database binding is the
same. For example, suppose we have an associative array $A$ and
database table binding $T$, finding all data that has a row key
between $a$ and $d$ is denoted as: $A(a:d, :)$ or $T(a:d, :)$
depending on whether information is being requested from the
associative array or database table. In both cases, the data returned
is in the form of an associative array.

In order to connect to different database
engines, D4M uses various connectors (either existing or custom built) to
connect to popular databases. As an example, the D4M-mySQL connection
is done by calling the Java Database Connector (JDBC) from D4M. While the current
implementation of D4M has a limited set of backend engines that are
supported, this number is increasing. 

A typical user workflow to develop an analytic on a large dataset will
be as follows. First, the user makes use of a schema to convert
their raw dataset (often in JSON, TSV, CSV format) into an associative
array. The user can then read a one or more associative arrays into
memory and develop the desired analytic. The user can verify correctness of analytic
with alternate pieces of the larger dataset. The user can then insert the full
dataset, converted to associative array format, into a database engine (or
set of databases). The user can then query for data which results in
an associative array that can be used directly in the analytic
developed.

One of the challenges in working with numerous backend databases is in
developing a uniform syntax and data format to put queries in the
context of a particular database or to cast information from one to
another in order to perform cross-DB analytics.

The Context operation is to provide explicit control of the backend
database or storage engine. The Cast operator is to move data between
storage and database engines.

In D4M, the context operation is done by using the DBserver command
which returns a DB object that contains information about the specific
database being connected to. Thus, when performing a query on a
backend database, the DB operator will use the correct context and
connector to perform the required query. The DBserver function in the
server binding returns an object to a DB that contains the host,
instance, and authentication information. 

%Context
\begin{lstlisting}[frame=single]
DB = DBserver(host,type,instanceName,user,pass)

Inputs: 
  host = database host name
  type = type of database  
  instanceName = database instance name
  username = username in database
  password = password associated with username

Outputs:
  DB = database object with a binding to specific DB
\end{lstlisting}

Once a DB object is created, one can perform database specific
operations such as \emph{ls} or create a binding to a specific table
in the database. If the requested table does not exist, a table with
that name will be created in the database. Binding to a table provides
functionality such as querying and inserting data.

\begin{lstlisting}[frame=single]
A = T(rows,cols)

Inputs:
  T = database table
  rows = row keys to select
  cols = column keys to select

Outputs:
 A = associative array of all non-empty row/columns
\end{lstlisting}

The following example describes how one can connect to a database and
return all the information in a particular table in the database.

\begin{lstlisting}[frame=single]
DB = DBserver('host,'type', db_name','user','pass')

table_list=ls(DB); % returns all tables in DB

T=DB('tab_name'); % Table binding to tab_name

A=T(:,:); % Entries of tab_name put in assoc array

\end{lstlisting}

In D4M, associative arrays can be used as the interface to cast information from one
database to another. Consider the following example of casting
data from mySQL to Apache Accumulo (a noSQL database). Of course, it
is up to the user to ensure that data can be cast from one database
to another (for example, certain databases may not support certain
datatypes or schemas). The following example describes how one could
cast data from mySQL to a noSQL database such as Apache Accumulo via
associative arrays.

\begin{lstlisting}[frame=single]
DBsql=DBserver('host,'mysql', 'sql_dbname','u','p');

DBnosql=DBserver('host','nosql','dbname','u','p');

T=DBsql('tabname'); % Tabname in sql_dbname

Asql=T(:,:); % Entries of tabname into Asql

Tnosql=DBnosql('tabname'); % Tabname dbname

put(Tnosql, Asql); %Insert into tabname in dbname

Anosql=Tnosql(:,:);%Entries of tabname into Anosql

\end{lstlisting}

One of the important aspects of D4M is the ability to easily add new
database engines via an API exposed by the database developer. In the
next section, we will discuss how a popular NewSQL database SciDB was
added to D4M. A thorough description of the D4M-Accumulo binding can
be found in~\cite{kepner2013d4m}.

\section{The SciDB-D4M Connection}
\label{scidb}

SciDB is a parallel database designed for multidimensional data
management with support for a variety of in-database computation and
analytics. SciDB has the ability to connect to a variety of client side
tools such as R, Python or a Web Browser. The SciDB coordinator is
responsible for moving data across back end data
storage~\cite{brown2010overview}. Connection to a SciDB server is
mediated via the coordinator. Other instances in the SciDB cluster are
referred to as worker nodes. SciDB represents data as multidimensional
arrays which are defined by specifying dimensions and
attributes. Dimensions are 64-bit integers, and attributes can be one
of many supported SciDB datatypes. SciDB supports a variety of
connection mechanisms such as JDBC or a SHIM. 

\begin{figure}
\centerline{
\includegraphics[width=2.8in]{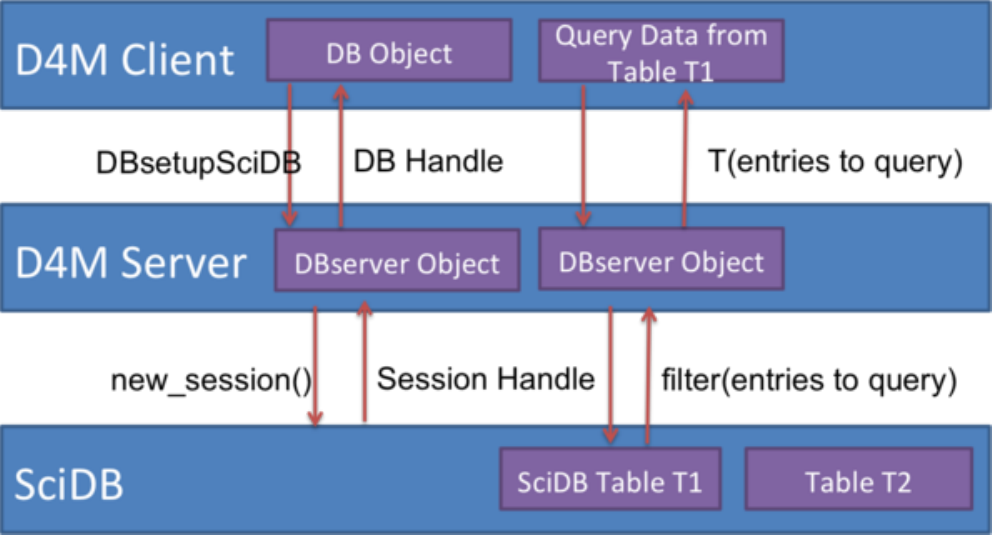}
}
\caption{D4M-SciDB Dataflow. A D4M operation will create a session
  with the SciDB coordinator node which will return information
  through D4M for requested data.}
\label{scidb-d4m}
\end{figure}

A SHIM is a small library that is capable of intercepting API calls
and translating them in to the underlying system API. In SciDB, the
Shim is a basic SciDB client that exposes SciDB functionality via a
HTTP interface~\cite{scidbshim}. The D4M-SciDB connection is built
using the SciDB SHIM. Specifically, given an operation on SciDB table, D4M
will convert this operation into a query that is supported by the
SciDB SHIM and pass it to the coordinator node. The coordinator node
will then perform the requested operation and return data back to D4M
via the established SHIM connection. As described in
Figure~\ref{scidb-d4m}, when a user calls a SciDB context
function, D4M will automatically translate the query into an operation
supported by the SciDB SHIM. When connecting to a SciDB table, a user will first call the \emph{DBserver}
operation that will authenticate and connect to SciDB via the
SHIM. This will return a token that is held in an object returned by
\emph{DBserver}. To establish a connection with an existing table in
SciDB, one can issue the D4M \emph{DBtable} command, that takes as an
argument the object returned by \emph{DBserver} and the required dimensions and
attributes. Currently, a number of D4M server binding commands are supported to
directly interface with the SciDB table. For example, \emph{nnz}, will
return the number of non-zero entries in a table. In any of these API
examples, the command issues the query to the backend database using
the context of the DB command.  Consider the example of inserting an associative
array into SciDB. The user will create an associative array and table
binding as described in Section~\ref{d4m}. The user can use the D4M
\emph{put} command which converts the associative
array into a datatype supported by SciDB and ingests this converted data to the
SciDB coordinator node. The dataflow is described in Figure~\ref{scidb-d4m}.

After
optimization, inserts are done in 128 MB batches and using the parallel
CSV loader. Once data is in SciDB, the standard D4M API can be used to
pull data back. For example, if the table binding is held in the
object \emph{T}, \emph{T(:,:)} returns all the elements in the table,
and \emph{T(row1:rowN, :)} returns the elements within the row range
\emph{row1:rowN}.

\subsection{D4M-SciDB Performance}

In order to benchmark SciDB, data was generated using a random graph
generator from the Graph500
benchmark~\cite{murphy2010introducing}. The Graph500 scalable data
generator that can efficiently generate power-law graphs that represent common graphs such as those generated from
social media datasets. The number of vertices and edges in the graph
are set using a positive integer called the SCALE parameter.  Given a SCALE parameter,
the number of vertices, N, and the number of edges, M, are then
computed as $N=2^{SCALE}$, and $M= 8N$.

\begin{figure}
\centerline{
\includegraphics[width=3.4in]{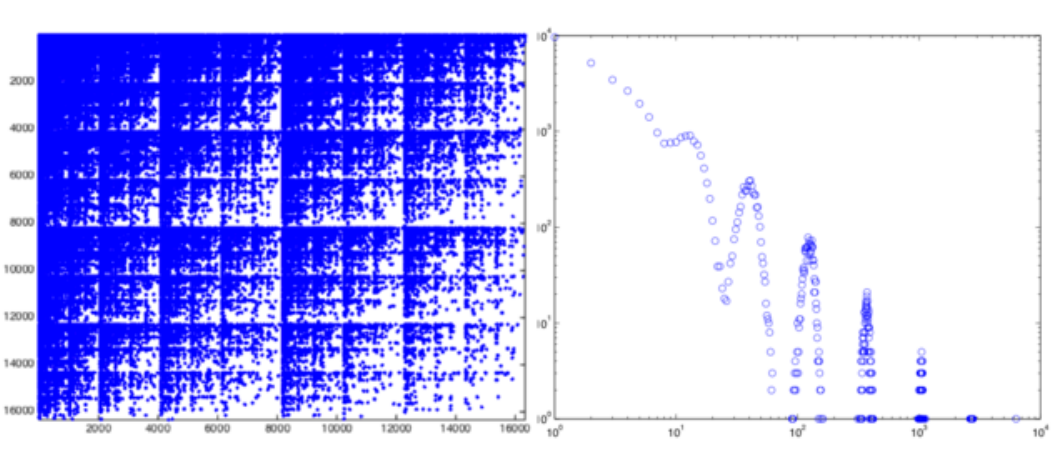}
}
\caption{Kronecker Graph Generator. The figure on the left represents
  the adjacency matrix. A connection between two vertices is denoted
  by a blue dot. The figure on the right shows the degree distribution
of the graph on the left.}
\label{kronecker}
\end{figure}

For example, if SCALE = 14, then N = 16384 and M = 131072.   The
Graph500 generator uses a recursive matrix
algorithm~\cite{chakrabarti2004r} to generate a set of starting
vertices and ending vertices corresponding to edges in a graph.  This graph is then be represented
as a large $N \times N$ sparse matrix $A$, where $A(i,j) = 1$ indicates an edge
from vertex $i$ to vertex $j$, often called the adjacency matrix.  As an example, consider
Figure~\ref{kronecker}, which shows the adjacency
matrix and distribution of degrees for a SCALE 14 graph generated
using the Kronecker graph generator. The degree of a vertex is the
number of edges incident to a vertex. For a power law graph, we expect
to see an exponential increase when looking at the number of nodes
with particular degrees (i.e, few nodes will have a high degree,
and many nodes will have a low degree).

SciDB is a highly scalable database and is capable of connecting with
multiple clients at once. In order to test the scalability of SciDB, we
use pMATLAB~\cite{bliss2007pmatlab} in addition to D4M to insert data
from multiple clients simultaneously. In order to overcome a SciDB
bottleneck that applies a table lock when data is being written to a
table, we use D4M to create multiple tables based on the total number
of ingestors. For example, if there are four simultaneous ingestors,
we create 4 tables into which each ingestor will simultaneously insert. The resulting tables can be merged after the ingest using D4M if desired.

SciDB was launched using the MIT SuperCloud~\cite{reuther2013llsupercloud} architecture through the
database hosting system. For the purpose of benchmarking SciDB on a
single node, instances were launched on a system with Intel Xeon E5
processors with 16 cores and 64GB of RAM. SciDB coordinator and worker
nodes were located on the same physical node.

\begin{figure}
\centerline{
\includegraphics[width=3.1in]{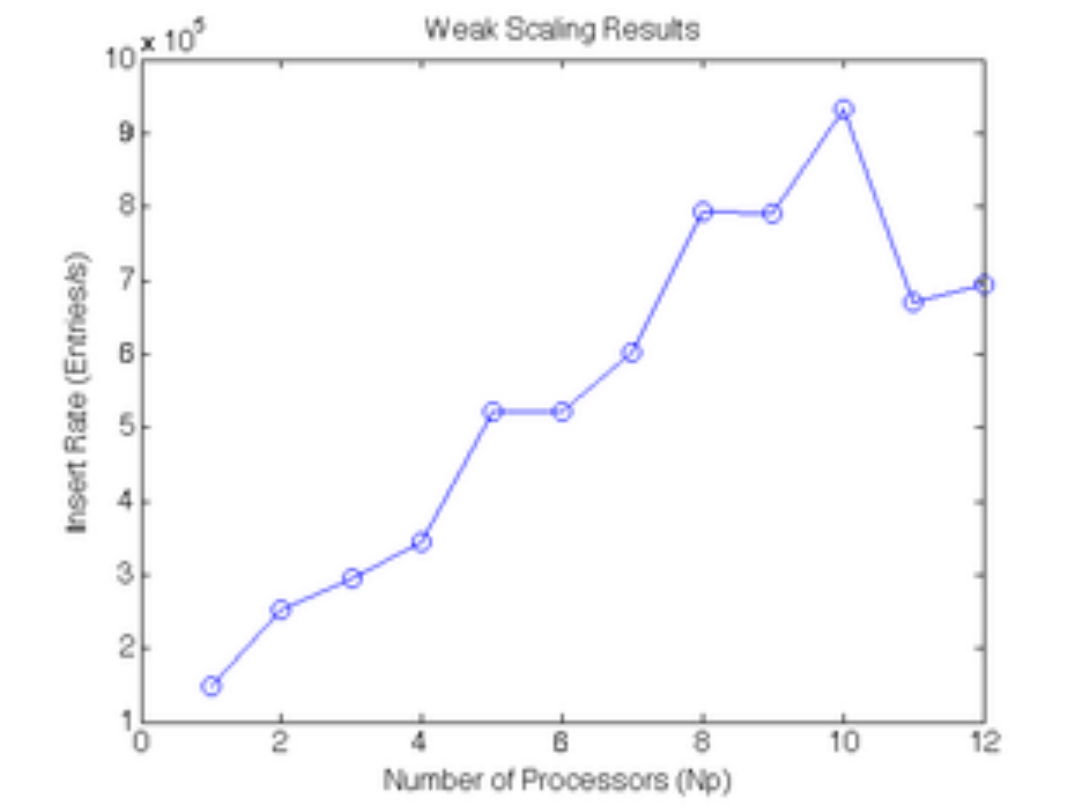}
}
\caption{Weak scaling of D4M-SciDB insert performance for problem
  size that varies with number of processors.}
\label{weakscaling}
\end{figure}

Weak scaling is a measure of the time taken for a single processing
element to solve a specific problem or fixed problem size per processor. In Figure~\ref{weakscaling}, we
describe the performance of SciDB in inserting 
Kronecker Graph whose SCALE varies with the number of processors into SciDB using D4M. The maximum performance (insert rate) was
observed at 10 processors.

\begin{figure}
\centerline{
\includegraphics[width=3.1in]{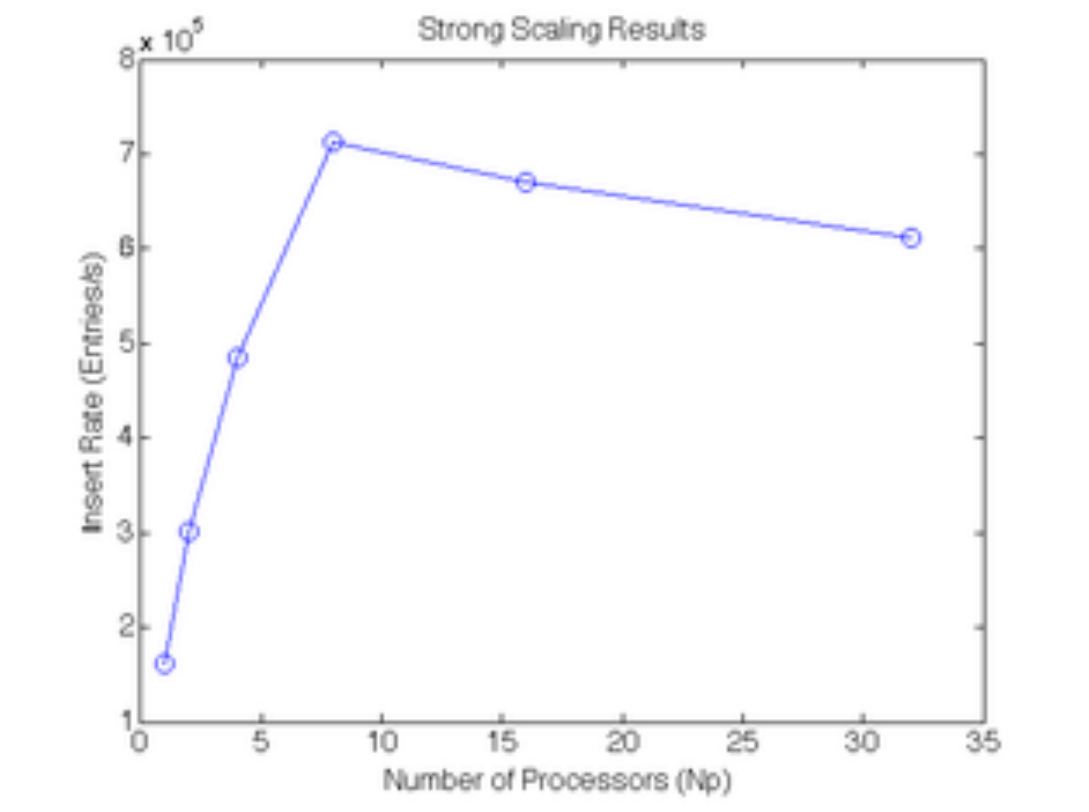}
}
\caption{Strong scaling of D4M-SciDB insert performance for varying
  number of processes for fixed SCALE=19 problem size. The y-axis represents the
insert rate.}
\label{strongscaling}
\end{figure}

Strong scaling is a measure of the time taken for solving a fixed
total problem size. Figure~\ref{strongscaling} describes the results varying
the number of inserters for a fixed SCALE 19 Kronecker Graph. The
maximum performance (insert rate) was observed to be at 8 processors.

\section{Medical Big Data Processing with Big Data}
\label{medical}

Medical big data is a common example used to justify the adage that ``one size
does not fit all'' for database and storage engines. Consider the popular MIMIC II dataset~\cite{MIMICII}. This dataset
consists of data collected from a variety of Intensive Care Units (ICU) at the
Beth Isreal Deaconess Hospital. The data contained in the MIMIC II
dataset was collected over seven years and contains data from a
variety of clinical and waveform sources. The clinical dataset contains the 
data collected from tens of thousands of individuals and consists of
information such as patient demographics, medications, interventions,
and text-based doctor or nurse notes. The waveform dataset contains
thousands of time series physiological signal recordings such as ECG
signals, arterial blood pressure, and other measurements of patient
vital signs. In order to support data extraction from these different
datasets, one option would be to attempt to organize all the information
into a single database engine. However, existing technologies would
prove to be cumbersome or inefficient for such a task. The next solution
is to store and index each of the individual components into a storage
or database engine that is the most efficient for a particular data
modality. While technically this solution may be the most efficient,
it makes application development difficult as researchers need to be
aware of underlying technologies and make development highly dependent
on changing technologies. D4M and associative arrays can be used to
provide developers (such as a medical researcher) with an abstraction
that hides such details in order to develop technology-agnostic
applications. As a part of the ISTC for Big Data, a prototype
application was developed that leverages different backend storage
engines. In this solution, the MIMIC II clinical data was placed in a relational database (MySQL), the text
notes were placed in Apache Accumulo, and the waveform data was placed
in SciDB using D4M.

\begin{figure*}[ht!]
\centerline{
\includegraphics[width=4.4in]{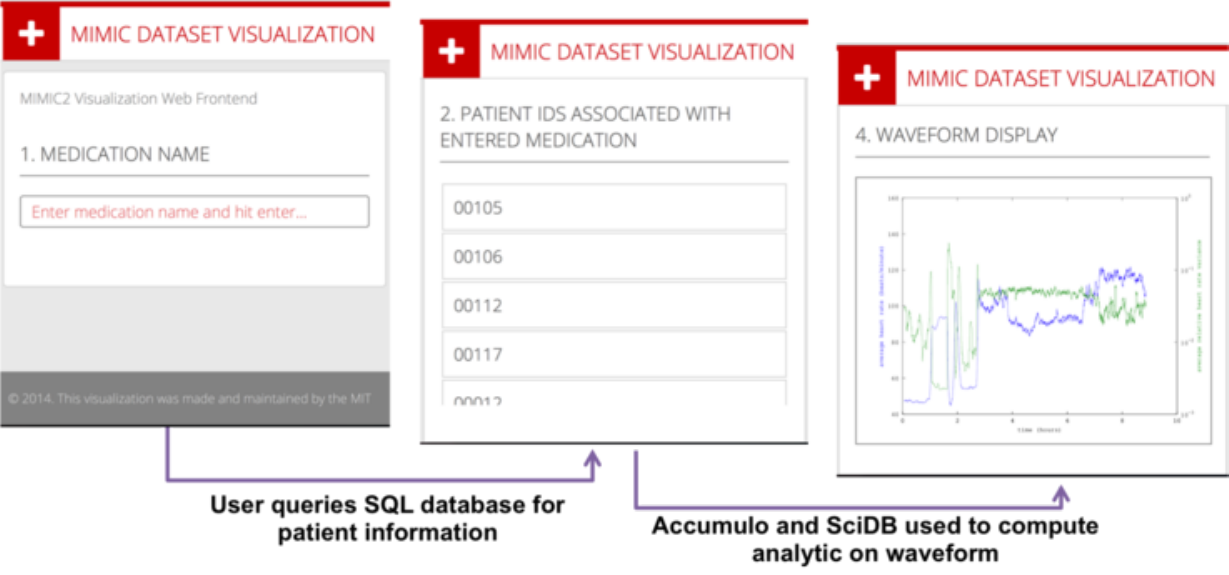}
}
\caption{Screen shots of MIMIC II Visualization that uses D4M and
  associative arrays for back end processing.}
\label{fi:mimicviz}
\end{figure*}

The prototype developed supports cross-database analytics such as:
``tell me about what happens to heart rate variance of patients who
have taken a particular medication.'' Naturally, such a query needs
information from
the clinical data contained in MySQL database, the patient
database contained in Accumulo and the waveform data
contained in SciDB. The sample query
provided is then be broken up into three distinct
queries where: 1) tell me which patients have taken a particular medication goes to
MySQL, 2) tell me which of these patients have heart
beat waveforms goes to Accumulo, and 3) show me what
happened to these patients heart rate variance goes to the waveform
database. At each of these sub-queries, associative arrays are
generated that can be used to move the results of one query to the next
database engine. In Figure~\ref{fi:mimicviz}, we show the web front end that
uses D4M and associative arrays to implement the query described
above. As an example of how D4M and associative arrays are used, 
querying the relational table results in an associative array
where the row keys represent the table name, and column keys represent the
patients who have taken Lisinopril. The resulting column keys are directly
passed into Accumulo to find all patients who have a
certain type of waveform where the rows contain the patient ID and
columns contain related waveform IDs. The resultant associative array can then be
passed directly into SciDB to extract the waveforms of interest and
perform the required analytic.

\section{Conclusions}
\label{conc}

D4M is a toolkit that supports a variety of database and storage
engine operations. Support for associative arrays can be used as a natural
interface between heterogenous database engines. Currently, D4M is
designed to work with a variety of engines such as SQL, Accumulo, and
SciDB. Currently, the number of context operations is limited; however,
D4M exposes these operations to the user with context specific
operations by allowing pass-through queries. Further, casting data
from one engine to another requires data to pass through the
client which may be a bottleneck for large scale data movement. 

In this paper, we described D4M and the relation between D4M,
associative arrays and databases. D4M can be used as a tool by
application developers to write applications agnostic of underlying
storage and database engines. Current research includes determining
how data can be cast directly between databases and increasing the
number of context agnostic D4M commands.

\section*{Acknowledgment}

The authors would like to thank the Intel Science and Technology
Center for Big Data and Paradigm4 for their
technical support in developing the D4M-SciDB connector.

\ifCLASSOPTIONcaptionsoff
  \newpage
\fi
\begin{spacing}{0.75}
\footnotesize
\bibliography{ieeehpec}
\end{spacing}
\vfill


\end{document}